\documentclass[runningheads]{llncs}
\usepackage{graphicx}
\usepackage{booktabs}
\usepackage{multirow}
\usepackage[table,xcdraw]{xcolor}
\usepackage{colortbl}
\begin{document}
\title{Exploring Teachers’ Perception of Artificial Intelligence: The Socio-emotional Deficiency as Opportunities and Challenges in Human-AI Complementarity in K-12 Education}

\titlerunning{Teachers' perception of AI}

\author{Soon-young Oh\inst{1}\orcidID{0000-0002-1507-6839} \and
Yongsu Ahn\inst{2}\orcidID{0000-0002-5797-5445}\thanks{Corresponding author}}
\institute{Michigan State University, East Lansing, MI 48823, USA \\ \email{ohsoon@msu.edu} \\
University of Pittsburgh, Pittsburgh, PA 15260, USA\\
\email{yongsu.ahn@pitt.edu}}
\maketitle              
\begin{abstract}
In schools, teachers play a multitude of roles, serving as educators, counselors, decision-makers, and members of the school community. With recent advances in artificial intelligence (AI), there is increasing discussion about how AI can assist, complement, and collaborate with teachers. To pave the way for better teacher-AI complementary relationships in schools, our study aims to expand the discourse on teacher-AI complementarity by seeking educators' perspectives on the potential strengths and limitations of AI across a spectrum of responsibilities. Through a mixed method using a survey with 100 elementary school teachers in South Korea and in-depth interviews with 12 teachers, our findings indicate that teachers anticipate AI's potential to complement human teachers by automating administrative tasks and enhancing personalized learning through advanced intelligence. Interestingly, the deficit of AI's socio-emotional capabilities has been perceived as both challenges and opportunities. Overall, our study demonstrates the nuanced perception of teachers and different levels of expectations over their roles, challenging the need for decisions about AI adoption tailored to educators' preferences and concerns.

\end{abstract}

\section{Introduction}
\label{sec:intro}

Teachers are composites that play an array of roles in schools. These roles encompass not only educational responsibilities such as teaching, guiding, and communicating with students and parents but also administrative duties ranging from document management, event coordination, and engagement in the decision-making process \cite{bidwell2013school}. The multifaceted nature of these roles necessitates that teachers possess a variety of cognitive abilities, for instance, as observers and mentors who provide guidance to students and as workers who efficiently process and organize information about students and the school community.

As artificial intelligence (AI) continues to advance and integrate into our society, discussions have emerged regarding its potential role in schools and its impact on teachers. Despite the overall benefits of introducing AI to schools, it is crucial to understand how teachers perceive the complementarity of AI. Such understanding -- whether they seek support in specific tasks or express resistance and skepticism -- can not only help identify potential barriers or opportunities to adopt AI capabilities but also tailor AI integration to their needs and concerns. However, in what \textit{capacities} do teachers anticipate AI acting as an assistant or collaborator in their duties, or automating and complementing their tasks to alleviate their burden? What \textit{tasks} do teachers believe AI can excel at or struggle with, shaping the dynamics of teacher-AI complementarity differently? While various studies explore teachers' perceptions of AI, the majority focus on classrooms \cite{kim2022teacher} or the educational roles of teachers \cite{woodruff2023perceptions,felix2020role}. As highlighted in existing work \cite{bidwell2013school}, teachers' roles and workload in the school scene span a broader spectrum of educational activities as well as administrative tasks, potentially leading to a variety of AI integration in education.

As an initial step toward fostering effective teacher-AI complementarity in schools, our research investigates teachers' expectations regarding AI capabilities across a range of eleven key teacher roles. Through a survey involving 100 teachers and in-depth interviews with 12 teachers, our findings demonstrate a diverse range of AI roles envisioned by teachers, spanning from document processing automation to roles as curriculum planners, decision-makers, and even leaders. The thematic analysis unveils a nuanced perspective: while teachers recognize AI's advanced intelligence, they highlight its deficiency in socio-emotional capabilities. Specifically, the AI's deficit of socio-emotional capabilities is perceived as opening up both opportunities and challenges. On one hand, teachers express concerns that AI’s inability to interpret nuanced student communication—both verbal and non-verbal—could hinder guidance and impede interpersonal growth by blurring the lines between human and AI interactions. On the other hand, the absence of emotion in AI is regarded as advantageous, positioning it as a fair and impartial entity capable of undertaking various tasks, ranging from task allocation to final decision-making within educational settings. 

\textbf{Contributions.} Overall, our study broadens the discussion on AI capabilities in education across a multitude of teachers’ roles. The contributions include: (1) \textbf{Teachers' task classification and cognitive mapping}: We develop a two-level classification of teachers' tasks in K-12 schools. These tasks are then mapped to cognitive abilities, allowing us to characterize them as a combination of different types of cognitive involvement. (2) \textbf{Techers' perception and imaginaries of future education}: We highlight the diverse array of roles spanning from automated document processing to tutoring and decision-making, underscoring AI's roles in future education beyond the current boundaries of AI advancements, including generative AI and large language models.

\section{Study Method}
\label{sec:study-method}

We employ a mixed method 1) to quantitatively derive the perceived AI complementarity using the survey method, and 2) to qualitatively investigate teachers’ thoughts and sense-making of AI’s potential opportunities and challenges in complementing their roles.

\textbf{Teacher task classification.} Our initial step involves filling a notable gap in the existing literature by creating a comprehensive classification of teacher tasks, drawing from classifications found in previous studies \cite{mintzberg1979structuring,bidwell2013school,hoy2008educational,kwon2010study}. We sought advice from six education experts with doctoral degrees in educational administration and technology. The classification from our foundational work (Fig. \ref{fig:result}) organizes teacher tasks into two levels, with the Level 1 (L1) covering broad domains, specifically educational and administrative tasks, and the Level 2 (L2) specifying a range of eleven sub-tasks, capturing both overarching job domains and the intricate details of individual task elements.

\textbf{Survey design.} To quantitatively measure teachers’ perception of AI’s roles, we utilized the Analytic Hierarchy Process (AHP), a method for deriving priorities through pairwise comparisons among factors within a decision-making hierarchy \cite{saaty1977scaling}. In this survey design, item pairs are presented for pairwise comparison, asking participants to score them on a scale ranging from -9 to 9. The lowest score indicates a strong preference for the left item over the right, while the highest score indicates the opposite preference. The AHP analysis results in per-task weights as normalized relative preferences summing up to 1. We refer to these weights as the Perceived AI Complementarity Score (PCS), the degree to which a task is perceived as more suitable for AI to complement human teachers.

\textbf{Recruitment and sampling.} We recruited 100 elementary school teachers in South Korea over two months, from August to September 2020. Snowball sampling was employed to construct a nationwide sample of teachers from various schools. We ensured the quality of all responses through two rounds of consistency ratio (CR) checks, maintaining a threshold of 0.1. Subsequently, 12 participants were selected for one-on-one, semi-structured interviews to explore detailed rationales behind their responses. Interviewees were purposefully chosen to represent diverse levels of experience and backgrounds. This included a balanced mix of regular teachers (50\%) and head teachers (50\%), and years of experience distributed as follows: $\leq$ 5 years (16.7\%), 6-10 years (25\%), 11-20 years (25\%), and $>$ 20 years (33.3\%).

\textbf{Task-ability association.} With teachers’ perception of tasks based on the survey responses, we provide a perspective of viewing tasks along the axis of cognitive abilities with the following question: What cognitive abilities do teachers perceive AI as being able (or not able) to complement? To conduct the analysis, we follow three analytical steps: 1) Task-ability mapping (Fig. \ref{fig:result}B): We mapped the relationship between eleven teacher tasks in our classification and fourteen cognitive abilities defined in \cite{tolan2021measuring}, which were derived from AI, animal, and psychological studies. This mapping involved annotation tasks on whether each task required specific cognitive abilities, represented in Fig. \ref{fig:result}B by green (indicating required) and gray (indicating not required). 2) Task group identification (Fig. \ref{fig:result}C): Utilizing the binary mapping between tasks and abilities, we performed a clustering analysis using the k-means method (k=4) to identify task groups of eleven teacher tasks based on their similarity in cognitive abilities and computed aggregate PCS scores. This allows us to investigate which cognitive abilities involved in tasks are perceived as having more AI complementarity.

\textbf{Interview procedure and analysis.} The one-on-one interview sessions were designed to allow participants to share their detailed thoughts on the complementarity between teachers and AI while answering a predetermined set of questions. The questions include: 1) Can you elaborate on your response in the survey regarding AI's complementarity in each task? 2) What opportunities and challenges do AI face in teacher-AI collaboration? 3) To what extent can AI complement human teachers in each task, and will AI assist, complement, or replace the role of humans?

To capture the essence of participants’ thoughts, we conducted a thematic analysis of the transcribed audio recordings from the interviews. Two coders, each with expertise in AI and education, transcribed and analyzed the interview, focusing mainly on two points: 1) What specific tasks do teachers perceive AI can or cannot perform better? 2) For those tasks, what opportunities or values and challenges or adversities do teachers perceive AI may encounter?

\section{Analysis Results} 
\label{sec:result}

\subsection{Teachers' perception: Administrative affairs perceived as the most viable tasks for AI}
\label{sec:quant}

The PCS scores derived from the AHP analysis (Fig. \ref{fig:result}A) in the L1 tasks indicate a preference for AI's involvement in school administration (sum of PCS in L2 tasks: 0.820) over educational tasks (0.180). In the L2 tasks, four sub-tasks in school administration emerged as the top priority, ranking from first to fourth among all tasks. Specifically, administrative affairs were identified as the highest priority (PCS: 0.437), followed by policy administration (PCS: 0.172), educational administration (PCS: 0.138), and external relations (PCS: 0.109), underscoring a belief in AI's greater efficacy in administrative roles.
\subsection{Task-ability mapping: Socio-emotional capabilities perceived as low AI proficiency}
\label{sec:mapping}

\begin{figure}
    \includegraphics[width=\textwidth]{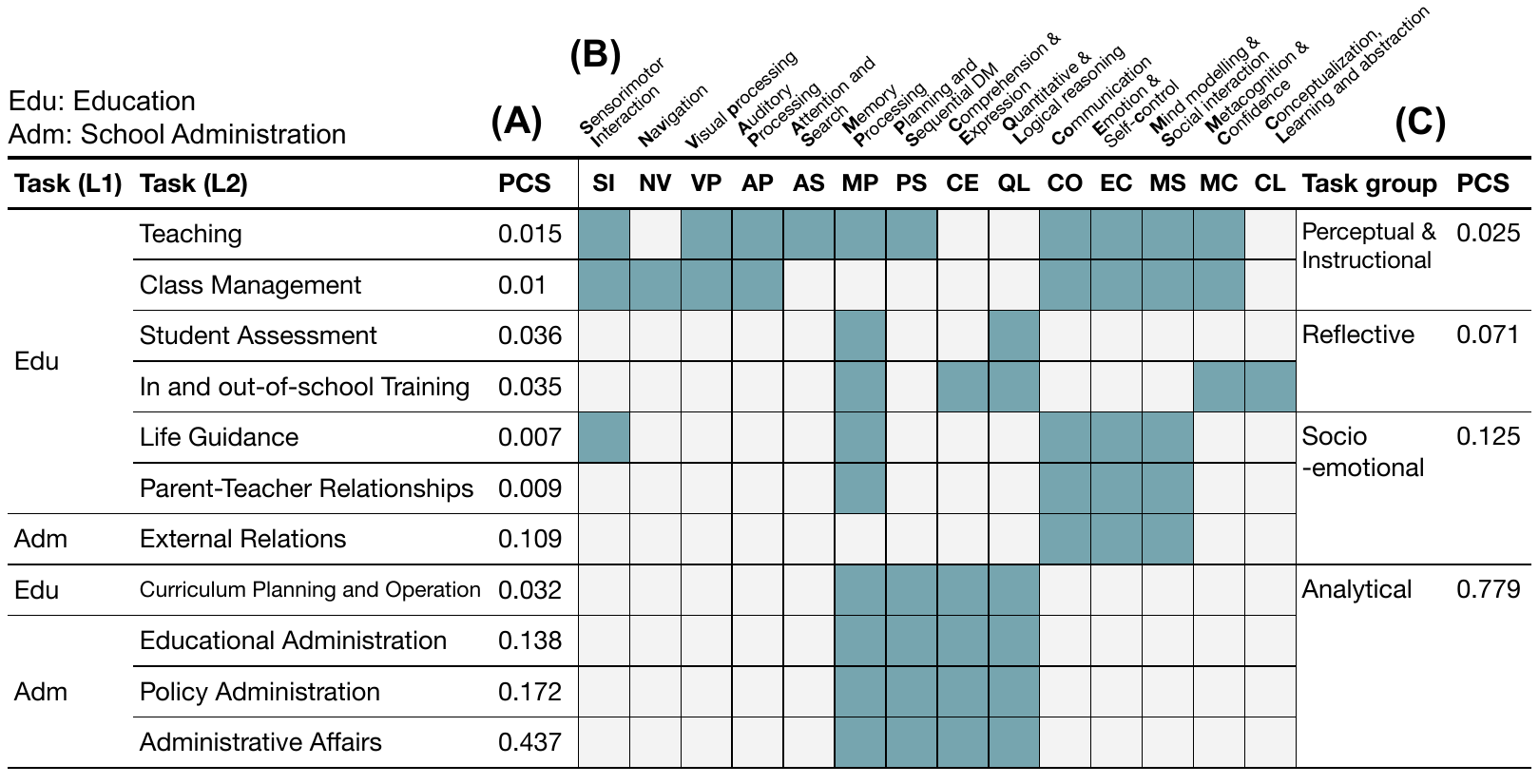}
    \vspace{-2em}
    \caption{\label{fig:result}
    \textbf{The overview of quantitative analysis and results.} (A) Task-wise perceived AI complementarity scores (PCS) from the survey using AHP method: Higher scores indicates AI being perceived as capable of complementing humans. (B) Task-ability mapping between tasks and abilities. (C) Task groups identified from clustering the tasks using the mapping in (B).}
    \vspace{-1.5em}
\end{figure}

As presented in Fig. \ref{fig:result}C, we identified four different groups of tasks named Perceptual \& Instructional, Reflective, Socio-emotional, and Analytical based on the k-means method. By averaging the PCS in each task group, we found that AIs are least perceived as complementing human teachers in teaching and class management that involve perceptual and instructional capabilities. The socio-emotional task group was also perceived as having low complementary. Especially, two of the specific tasks, life guidance and parent-teacher relationships as individual tasks, obtained the lowest perceived complementarity score, showing the challenging nature of socio-emotional capabilities as AI-complementary tasks. On the other hand, analytical tasks were dominantly perceived as the most capable complementarity duties AI can take on.
\vspace{-0.5em}
\subsection{Opportunities and challenges of AI: Advanced intelligence and socio-emotional deficiency}
\label{sec:qualitative}

The thematic analysis of in-depth semi-structured interviews highlights two focal themes of teachers’ perception on characteristics of AI capabilities: (1) advanced intelligence and (2) lack of socio-emotional capabilities. We show that the 68 meaningful comments mentioned by 12 teacher interviewees (referred to as T1-12) (Table \ref{tbl:qual}), as opportunities (marked as {\color{blue}up-arrows}) and challenges (marked as {\color{red}down-arrows}) of teacher-AI complementarity, largely fall into either of these two characteristics of AI, particularly highlighting AI's socio-emotional deficiency perceived as both its strengths and limitations in the educational contexts.

\vspace{-1em}
\subsubsection{Advanced intelligence in personalized learning and automation.}
First, teachers highlighted various aspects of teacher tasks where AI could contribute to advancing their educational and administrative tasks. Notably, many comments emphasized the potential role of AI in personalized learning and student management. Envisioning AI's advanced intelligence, teachers anticipated capabilities such as \textit{“tailoring learning materials to individual student’s academic status”} or \textit{“providing daily check-ups and feedback.”}. As T2 noted, this could significantly benefit teachers in academic management, \textit{“Teachers and parents often struggle to assess students' academic progress consistently, such as solving math problems, on a daily basis.”}. Several teachers (T2, 8, 9, 10) expected that such capabilities in personalization, when provided at a large scale for hundreds of individual students, can potentially help address learning disparities, especially to minority and low-achieving students.

\vspace{-1em}
\begin{table}[]
\begin{tabular}{@{}lll@{}}
\toprule
\textbf{\begin{tabular}[c]{@{}l@{}}Task \\ \newline \\ \newline \end{tabular}}                                                               & \textbf{\begin{tabular}[c]{@{}l@{}}Advanced intelligence\\ perceived as \textcolor{blue}{opportunities} \\ \newline \end{tabular}}                          & \textbf{\begin{tabular}[c]{@{}l@{}}Socio-emotional deficiency\\ perceived as \textcolor{blue}{opportunities}\\ and \textcolor{red}{challenges}\end{tabular}}                                                                                                                                                                                              \\ \midrule

\begin{tabular}[c]{@{}l@{}}Teaching \\ \newline \\ \newline \\ \newline \end{tabular}             & \begin{tabular}[c]{@{}l@{}}\textbf{\textcolor{blue}{(↑)}} Personalized learning \\ \textbf{\textcolor{blue}{(↑)}} 1-on-1 education \\ (Equity) \\ \newline \end{tabular} & \multirow{-2.5}{*}{\begin{tabular}[c]{@{}l@{}}\textbf{\textcolor{red}{(↓)}} Emotion/intent recognition\\ \textbf{\textcolor{red}{(↓)}} Social interaction\\ \textbf{\textcolor{red}{(↓)}} Non-verbal communication\\ \textbf{\textcolor{red}{(↓)}} Persuasion, Guidance\\  \textbf{\textcolor{blue}{(↑)}} Consistent communication\\ management (Fairness) \end{tabular}} \\ \cmidrule(r){1-2}
\begin{tabular}[c]{@{}l@{}} Life guidance \\ \newline \end{tabular}                                                              & \begin{tabular}[c]{@{}l@{}}\textbf{\textcolor{blue}{(↑)}} Personalized management \\ \newline \end{tabular}                                     &                                                                                                                                                                                                                                                                                 \\ \cmidrule(r){1-2}
Class management & - & \\ \cmidrule(r){1-2}
\begin{tabular}[c]{@{}l@{}}Parent-teacher\\ Relationships\end{tabular}      & \begin{tabular}[c]{@{}l@{}} - \\ \newline \end{tabular}                                                                                                                          &                                                                                                                                                                                                                                                                                                                        \\ \midrule
\begin{tabular}[c]{@{}l@{}}Student\\ Assessment \\ \newline \end{tabular}            & \begin{tabular}[c]{@{}l@{}}\textbf{\textcolor{blue}{(↑)}} Personalized feedback \\ \newline \\ \newline \end{tabular}                                                                                                  & \begin{tabular}[c]{@{}l@{}}\textbf{\textcolor{red}{(↓)}} Behavioral feedback \\ \textbf{\textcolor{blue}{(↑)}} Automated scoring \\ (Fairness, Efficiency)\end{tabular}                                                                                                                                                     \\ \midrule

\begin{tabular}[c]{@{}l@{}}In and out-of-school\\ training\end{tabular}    & \begin{tabular}[c]{@{}l@{}}\textbf{\textcolor{blue}{(↑)}} AI-powered instruction\\and personalized training\end{tabular}                               & \begin{tabular}[c]{@{}l@{}}- \\ \newline \end{tabular}                                                                                                                                                                                                                                                                                                                     \\ \midrule
\begin{tabular}[c]{@{}l@{}}Curriculum planning\\ and operation\end{tabular} & \begin{tabular}[c]{@{}l@{}}- \\ \newline \end{tabular}                                                                                                                          & \begin{tabular}[c]{@{}l@{}}\textbf{\textcolor{blue}{(↑)}} Data-driven planning \\ (Fairness)\end{tabular}                                                                                                                                                                                                                          \\ \midrule
\begin{tabular}[c]{@{}l@{}}Educational\\ administration\end{tabular}        & \multirow{3}{*}{\begin{tabular}[c]{@{}l@{}}\textbf{\textcolor{blue}{(↑)}} Automation \\ (Efficiency) \\ \newline \\ \newline \\ \newline \end{tabular}}                                                                                & \multirow{3}{*}{\begin{tabular}[c]{@{}l@{}}\textbf{\textcolor{blue}{(↑)}} Data-driven decision making \\ and task assignment \\ (Fairness, Deauthorization) \\ \newline 
 \end{tabular}}                                                                                                                                                                          \\ \cmidrule(r){1-1}
Policy administration                                                       &                                                                                                                              &                                                                                                                                                                                                                                                                                                                        \\ \cmidrule(r){1-1}
Administrative affairs                                                      &                                                                                                                              &                                                                                                                                                                                                                                                                                                                        \\ \midrule
\begin{tabular}[c]{@{}l@{}}External relations \\ \newline \end{tabular}                                                             & \begin{tabular}[c]{@{}l@{}}- \\ \newline \end{tabular}                                                                                                                            & \begin{tabular}[c]{@{}l@{}}\textbf{\textcolor{red}{(↓)}} Negotiation, Interpersonal \\ relationship\end{tabular}                                                                                                                                                                                                                                 \\ \bottomrule
\end{tabular}
\label{tbl:qual}
\caption{\label{tbl:qual} Two core perceived characteristics of AI, advanced intelligence and socio-emotional deficiency, perceived as its opportunities (marked as {\color{blue}up-arrows}) and challenges ({\color{red}down-arrows}) in K-12 education.}
\end{table}
\vspace{-2em} 

Furthermore, the majority of administrative tasks were perceived as significant opportunities with the help of automation. Teachers found that some tasks, such as document processing or budget planning (T7), are typically \textit{``fixed and standardized''} (T2, 10, 12) or \textit{``handled on an annual or monthly basis''} (T7). Educational administrations such as curriculum planning were also expected for AI to not only automate tasks, but also advance the planning to be more contextualized to each school environment based on learning a variety of data from both national and school levels (T9, 11).

\subsubsection{Lack of social-emotional capabilities as challenges in teaching and life guidance.}
Simultaneously, teachers tended to perceive AI as lacking socio-emotional capabilities, particularly when it comes to tasks that involve guiding students during teaching and class activities. Participants stated that such tasks often pose one of the greatest challenges even for human teachers, resulting from tricky interactions, as T9 commented, \textit{``During class, many students claim that they are paying attention while actually being distracted and engaged in unrelated activities, or pretending that they understood everything.''} Several teachers highlighted that such situations require the ability to recognize nuanced verbal and non-verbal communications and respond in a way that motivates and guides students to stay focused in class. During the interview, these complexities were often described as \textit{``there are too many variations''} (T1, 10), as \textit{``[it] needs a significant level of adaptability with experiences and careful observations''} (T11). Most of the teachers in the study were skeptical about whether AI can truly \textit{``grasp these subtle cues, relationships among students such as jealousy, and handle issues like parental complaints''} (T9), given that these interactions often have no obvious answers and involve a lack of available and unstructured data (T1).

Teachers found it especially challenging in elementary school, where students before the age of 18 ego development in human-human communication, \textit{``I'm worried that children, still navigating their ego development, may struggle to differentiate between human and AI interaction, potentially leading to confusion about their own identity.''} (T12), or \textit{``Because interpersonal relationships are crucial, children's feelings of isolation could intensify.''} (T1)

Additionally, teachers recognized AI's potential in facilitating connections with the local community and parents, with educators emphasizing the indispensability of human-to-human rapport. As one teacher stated, \textit{``Building relationships with the local community requires genuine human interaction''} (T9), and \textit{``[such tasks] should involve a political aspect within subtle dynamics of relationships with the local community using negotiation or political skills''} (T10).

\subsubsection{Lack of socio-emotional capabilities as opportunities for fair and non-authoritative decision-making.}
Despite AI’s perceived limitation on teaching and guiding due to AI’s socio-emotional deficiency, this was, at the same time, an opportunity for AI to step into schools as a fair and nonauthoritative agency at various levels of teachers’ tasks. Participants, as members of an educational organization, felt that \textit{``decisions in schools often get swayed by those with a louder voice, or who've taken on senior roles.''} (T4)  or experienced conflicts between them in task assignment, \textit{``When a human teacher handles it, there's often a lot of conflict about who's responsible for certain tasks in school management, like whether this is my job or yours.''} (T9) For instance, they suggested that AI could provide feedback such as, \textit{``[AI] could say like, ``in the past, making choices like this drew a lot of criticism.'' It could provide a basis for judgment.''} (T4)

T9 shared more radical imaginaries of AI roles as a decision maker and leader to foster equal and non-authoritative cultures. 

\vspace{-0.5em}
\begin{quote}
\textit{``Leveraging AI is about simply assigning it a leader role. So, if we prompt AI to take on the principal's role and let it allocate suitable roles to teachers, we could prevent conflicts and establish a more horizontal structure.''}
\end{quote}
\vspace{-0.5em}

Additionally, teachers, as graders and mentors, expressed a challenge in achieving fair evaluation during grading and providing feedback. T1 articulated this concern, saying, 

\vspace{-0.5em}
\begin{quote}
\textit{``In grading, there are situations where emotions come into play. I sometimes wish for partial credit when a student answers incorrectly, especially if they've usually been good. So, there are moments when you think, `Hmm, maybe they deserve some points here,' or the other way around. And that's where I see artificial intelligence might be really helping out, making evaluations more objective and fair.''}
\end{quote}
\vspace{-1em}
\section{Conclusion}
\label{sec:discussion}

This paper investigates teachers' perceptions of AI capabilities across various teacher tasks. Through survey data and in-depth interviews, we uncover a spectrum of opportunities and challenges associated with AI, characterized by its advanced intelligence yet socio-emotional limitations. Our study expands the discourse on the future of AI across a range of teacher tasks, with an in-depth discussion of teachers’ perspectives on AI capabilities beyond recent developments in generative AI.


\begin{thebibliography}{8}
\bibitem{woodruff2023perceptions}Woodruff, K., Hutson, J. \& Arnone, K. Perceptions and barriers to adopting artificial intelligence in K-12 education: A survey of educators in fifty states. (2023)

\bibitem{felix2020role}Felix, C. The role of the teacher and AI in education. {\em International Perspectives On The Role Of Technology In Humanizing Higher Education}. pp. 33-48 (2020)

\bibitem{kim2022teacher}Kim, N. \& Kim, M. Teacher’s perceptions of using an artificial intelligence-based educational tool for scientific writing. {\em Frontiers In Education}. \textbf{7} pp. 142 (2022)

\bibitem{celik2022promises}Celik, I., Dindar, M., Muukkonen, H. \& Järvelä, S. The promises and challenges of artificial intelligence for teachers: A systematic review of research. {\em TechTrends}. \textbf{66}, 616-630 (2022)


\bibitem{mintzberg1979structuring}Mintzberg, H. The structuring of organizations Prentice-Hall. {\em Englewood Cliffs, NJ}. (1979)

\bibitem{bidwell2013school}Bidwell, C. The school as a formal organization. {\em Handbook Of Organizations (RLE: Organizations)}. pp. 972-1022 (2013)

\bibitem{hoy2008educational}Hoy, W. \& Miskel, C. Educational administration: Theory, research, and practice. {\em (No Title)}. (2008)

\bibitem{kwon2010study}Kwon, H. A study on how to establish job standards for elementary school teachers. {\em The Journal Of Korean Teacher Education}. \textbf{27}, 191-214 (2010)

\bibitem{saaty1977scaling}Saaty, T. A scaling method for priorities in hierarchical structures. {\em Journal Of Mathematical Psychology}. \textbf{15}, 234-281 (1977)

\bibitem{tolan2021measuring}Tolan, S., Pesole, A., Martinez-Plumed, F., Fernandez-Macias, E., Hernandez-Orallo, J. \& Gomez, E. Measuring the occupational impact of ai: tasks, cognitive abilities and ai benchmarks. {\em Journal Of Artificial Intelligence Research}. \textbf{71} pp. 191-236 (2021)

\end{thebibliography}
\end{document}